\documentclass[prl,twocolumn,superscriptaddress]{revtex4}
\usepackage{graphicx}
\usepackage{dcolumn}
\usepackage{amssymb}
\usepackage{bm}
\usepackage{pifont}
\usepackage{epsfig}
\usepackage{psfrag}
\usepackage{epstopdf}
\usepackage{wasysym}
\usepackage[usenames]{color}
\usepackage[symbol]{footmisc}
\def\correspondingauthors{\footnote{vi21@georgetown.edu}}
\def\correspondingauthor{\footnote{dlb76@georgetown.edu}}
\def\correspondingauth{\footnote{urbachj@georgetown.edu}}

\begin{document}

\title{Localized Transient Jamming in Discontinuous Shear Thickening }

\author{Vikram Rathee \correspondingauthors{} }
\affiliation{Department of Physics, Georgetown University, Washington, DC 20057; and Institute for Soft Matter Synthesis and Metrology, Georgetown University,
Washington, DC 20057 }

\author{Daniel L. Blair \correspondingauthor{}}
\affiliation{Department of Physics, Georgetown University, Washington, DC 20057; and Institute for Soft Matter Synthesis and Metrology, Georgetown University,
Washington, DC 20057 }

\author{Jeffrey S. Urbach \correspondingauth{}}
\affiliation{Department of Physics, Georgetown University, Washington, DC 20057; and Institute for Soft Matter Synthesis and Metrology, Georgetown University,
Washington, DC 20057 }


\begin{abstract}
We report direct measurements of spatially resolved surface stresses over the entire surface of a dense suspension during discontinuous shear thickening (DST) using Boundary Stress Microscopy (BSM) in a parallel-plate rheometer.
 We find that
large fluctuations in the bulk rheological response at the onset of
DST are the result of localized transitions to a state with very high
stress, consistent with a fully jammed solid that makes direct contact with the shearing boundaries.  That jammed solid like phase (SLP) is
rapidly fractured, producing two separate SLPs that propagate in opposite directions.  By comparing the speed of propagation of the SLPs with the motion of the confining plates, we deduce that one  
remains in contact with the bottom boundary, and another  remains
in contact with the top.  These regions grow, bifurcate, and
eventually interact and decay in a complex manner that depends on the measurement conditions (constant shear rate vs constant stress).  
In constant applied stress mode,  BSM directly reveals dramatic stress fluctuations that are completely missed in standard bulk rheology.

 \end{abstract}

\maketitle

\section{Introduction}

An increase in viscosity, $\eta$, above a material dependent critical
shear stress is commonly observed in dense colloidal and granular
suspensions \cite{Barnes, Lootens, Wagner_1,  Jaeger_1, Bonn_1, Ovarlez,  Poon_1}. This increase can be
abrupt if the volume fraction ($\phi = V_{{\rm particle}}/V_{{\rm total}}$)
approaches the jamming fraction, $\phi_{J}$, and is known as discontinuous
shear thickening (DST), while at lower concentrations the increase is
gradual and termed  continuous shear thickening (CST) \cite{Barnes, Lootens, Wagner_1,  Jaeger_1, Bonn_1,  Ovarlez, Poon_1, Lootens_2, Wagner_2,  Bonn_2, Seto, Isa_1, Cates_1, Claus, John, Itai_1, vikram}.

CST can arise from the formation of particle clusters  due to hydrodynamic forces \cite {, Wagner_2, Melrose, Brady, Wagner_3}. However recent work has shown that solid contact and friction, together with the existence of a short-range repulsive force between particles, are likely to play a major role in shear thickening \cite{Seto, Mari14, Cates_1, Mari15}.
Wyart and Cates (WC) introduced a phenomenological model in
which the shear thickening arises from a transition from primarily
hydrodynamic interactions when the applied stress is substantially
below a critical stress, $\sigma \ll \sigma ^*$, to primarily
frictional interactions when $\sigma \gg \sigma ^*$ \cite{Cates_1}.
A variety of results from experiments and simulations provide support
for the WC model~\cite{Itai_1, John, Colin, Poon_1, Clavaud}, although many
questions remain.

One of those questions concerns spatiotemporal dynamics in the shear
thickening regime.  Temporal fluctuations in bulk viscosity have been
observed in macroscopic rheology in DST, and visual observations
suggest some associated spatial heterogeneities  \cite{Lootens, Poon_1}.  Similarly, numerical simulations have revealed
stress fluctuations at the particle level and at larger scales
\cite{Claus, Claus_2}.
A recent study using ultrasound imaging to measure velocity profiles
in sheared cornstarch suspensions in the DST regime revealed bands of varying shear rate and wall slip
that propagate along the vorticity direction and proliferate as the
stress is increased \cite{Manneville}.

Using a novel technique that we have termed boundary stress microscopy
(BSM, \cite{Rich}), we have recently shown that complex spatiotemporal
dynamics between high and low stress phases are actually present in
the CST regime \cite{vikram, Park}.  BSM revealed the existence of clearly
defined dynamic localized regions of substantially increased stress
that appear intermittently at stresses above the critical stress.
We interpreted those regions as
high-viscosity fluid phases, consistent with the high viscosity
(frictional) fluid branch predicted by the WC model. In the present
work, we extend the application of BSM to a suspension of higher
concentration, where the bulk rheological response is in the DST regime, and use a custom rheometer tool that
enables us to perform BSM on the entire sample surface.  We show that
large fluctuations in the bulk viscosity at the onset of
DST are the result of localized transitions to a state with very high
stress, consistent with the existence of localized regions of a fully jammed solid that make direct
frictional contact with the shearing boundaries.  That region rapidly bifurcates, producing two separate jammed regions, one that
remains in contact with the bottom boundary, and another that remains
in contact with the top boundary.  These regions grow, bifurcate further, and
eventually interact and decay in a manner that suggests a complex
coupling between heterogeneous stresses, non affine flow, and density
fluctuations.


\section{Materials and methods}
Elastic films of thickness of 53 $\pm 3 \mu$m  were deposited by spin coating PDMS (Sylgard 184; Dow Corning) and a curing
agent  on 40 mm diameter glass cover slides (Fisher Sci) that were cleaned thoroughly by plasma cleaning and rinsing
with ethanol and deionized water \cite{vikram}. The PDMS and curing agent were mixed and degassed
until there were no visible air bubbles.  We observed these air-cured films have an elastic modulus (GÕ) in the range 15-20 kPa, significantly higher than the 10 kPa modulus of PDMS gels of the same composition but cured in-situ on the rheometer.  After deposition of PDMS, the slides
were cured at 85 $^{o}$C for 45 min. After curing, the PDMS was
functionalized with 3-aminopropyl triethoxysilane (Fisher Sci) using
vapor deposition for $~$40 min. For imaging, carboxylate-modified
fluorescent spherical beads of radius 5 $\mu$m with
excitation/emission at 520/560 nm were attached to the PDMS
surface. Before attaching the beads to functionalized PDMS, the beads
were suspended in a solution containing PBS solution
(Thermo-Figher). The concentration of beads used was 0.006 $\%$
solids. A second PDMS film of thickness $~$ 5-7
$\mu$m was added by spin coating to avoid bead detachment under shear.
Suspensions were formulated with silica spheres of radius, a = 0.75
$\mu$m (Angstorm, Inc.) suspended in a glycerol water mixture (0.8
glycerol volume fraction).  Rheological measurements were
performed on a stress-controlled rheometer (Anton Paar MCR 301)
mounted on an inverted confocal (Leica SP5) microscope \cite{Sudeep}
using a home made parallel plate tool of diameter 5.2 mm diameter. The gap between rheometer plates was fixed at 0.2 mm.  A 1.6X
objective was used for imaging and produced a 5.6 X 5.6 mm$^{2}$ field
of view. 
Deformation fields were determined with particle image velocimetry (PIV) in ImageJ \cite{Tseng}. The surface stresses at the interface are calculated using an extended traction force technique and codes given in ref. \cite{Style}, which produces a spatial map of the stress in the plane of the surface, $\vec{\sigma} (\vec{r})$.  Measurement noise arises primarily from the resolution of the PIV technique.  The stress maps reported in Figs. \ref{42event}, \ref{45event} and \ref{50Paevent} and shown in the supplementary movies are the stress magnitude,  $|\vec{\sigma} (\vec{r})|$.  In order to compare with the stress reported by the rheometer (which is calculated based on the torque applied to the tool), we calculate the average boundary stress as $\langle \sigma_{BSM} \rangle = \int _0 ^R \sigma _{\theta} (\vec{r})  r dA$, where $R$ is the radius of the tool.  Stresses inferred from the rheometer were calculated using the measured torque (M) according to   $\sigma=  \frac{2 M}{\pi R^{3}}$. The shear rate  was calculated from the angular velocity ($\omega$) of the tool  as $ \dot{\gamma} =\frac{R}{h}\omega $, where $h$ is the gap between the plates.


\section{Results}
\begin{figure}
\includegraphics[width=0.4\textwidth]{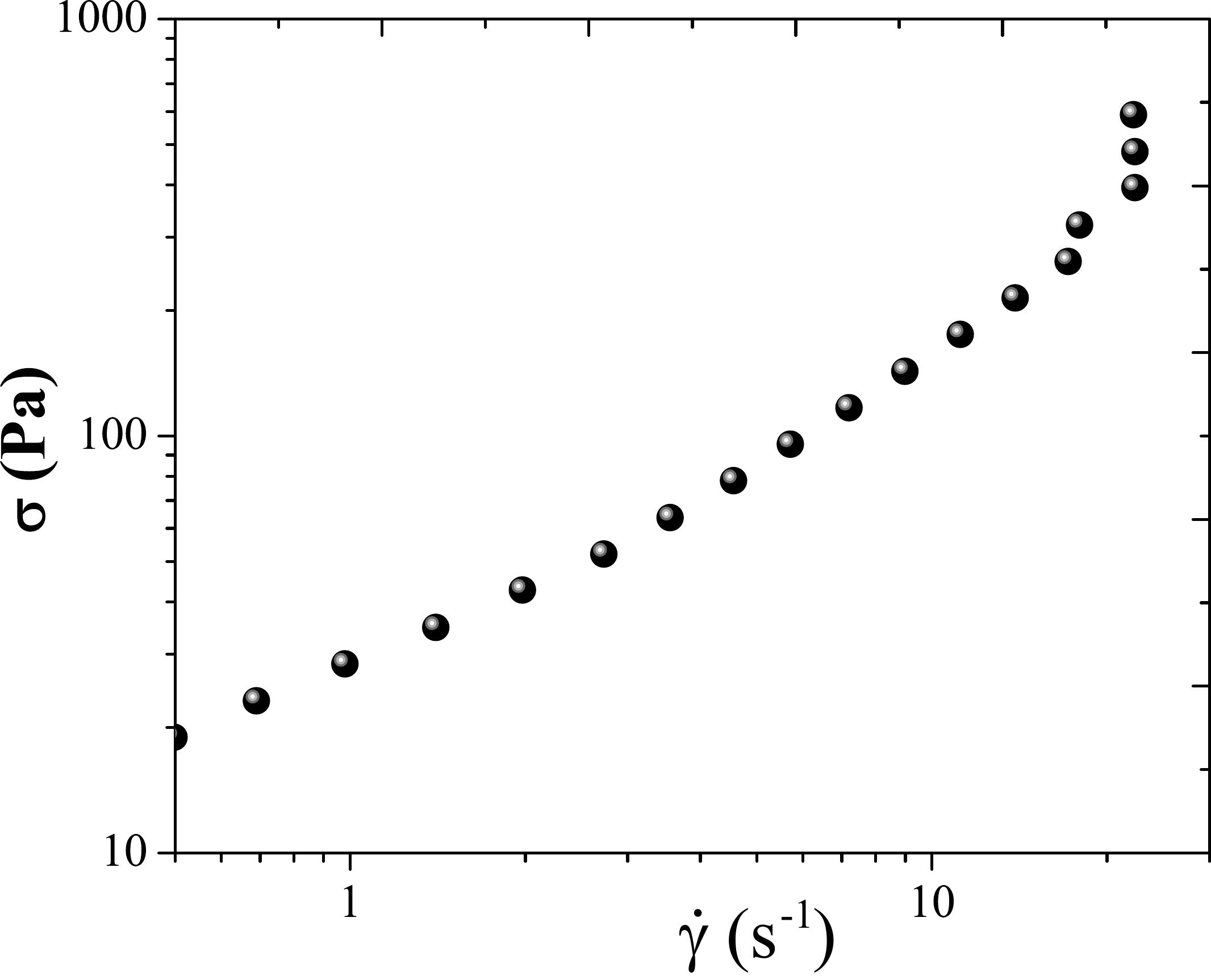}
\caption{ Stress vs shear rate  for a suspension with volume fraction ($\phi$) = 0.56  using a home made parallel plate rheometer tool of diameter 5.2 mm.}
\label{flowcurve}
\end{figure}

As described in the introduction, dense suspensions can show dramatic shear thickening above a critical shear stress.  Figure \ref{flowcurve} shows the flow curve for the suspension studied here ($\phi=0.56$). The average  viscosity, $\eta$, as a function of the stepwise increasing applied shear stress $\sigma$,  measured with the custom small (5.2 mm diameter) parallel plate rheometer tool,   indicates that at this volume fraction  the suspension is in the discontinuous shear thickening (DST) regime.  

\subsection{Constant Applied Shear Rate}
\begin{figure*}
\includegraphics[width=1\textwidth]{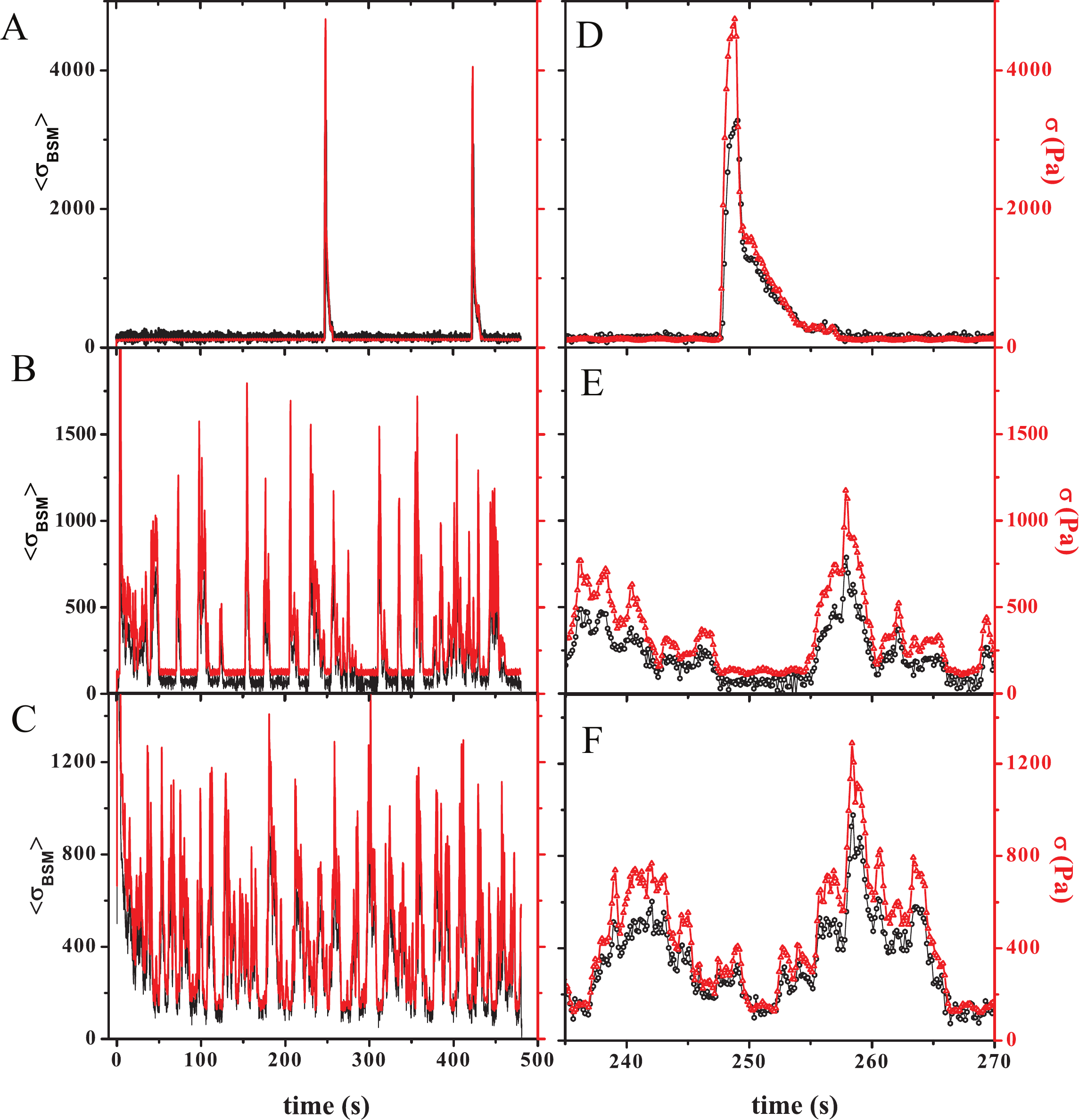}
\caption{ Temporal evolution of stress ($\sigma$, red) and average
  boundary stress ($\sigma_{{\rm BSM}}$, black) using boundary stress measurements (BSM) for different applied shear rate A,D) 27.3 s$^{-1}$, B,E) 32.5 s$^{-1}$ and C,F) 39 s$^{-1}$ for $\phi$ = 0.56. }
\label{stressvstime}
\end{figure*}

In the constant shear rate measurement mode, the rheometer controls the applied torque (and therefore average shear stress) in order to maintain a constant rotation rate (and therefore constant local shear rate, albeit one that varies with radius in the parallel plate geometry).
At relatively small edge shear rates, $\dot{\gamma} \leq$ 26 s$^{-1}$, we observe very small temporal oscillations around an average $\sigma$, possibly due to imperfections in the tool (data not shown). When a critical shear rate  of $\dot{\gamma}_{c} \approx 26.5$  $s^{-1}$ is reached, the rheometer (average) $\sigma$ reveals large intermittent fluctuations that are clearly higher than the background  of 30 Pa (Fig. \ref{stressvstime}A, red curves). As $\dot{\gamma}$ is increased further, the spikes in $\sigma$  become more frequent (Fig. \ref{stressvstime}B,C).  Similar behavior is observed with the same measurement geometry but without the elastic PDMS layer (Supplementary Fig. 1), suggesting that the fluctuations do not arise from the compliance of the elastic layer.

In addition to the average stress $\sigma$ reported by the rheometer,
we can measure the average boundary stress, $\langle \sigma_{{\rm
    BSM}}\rangle$, calculated directly from the deformation on the
elastic substrate.  Below $\dot{\gamma}_c, \langle \sigma_{{\rm
    BSM}}\rangle$ remains approximately constant, and barely above the measurement noise (data not shown). Above $\dot{\gamma}_c$, the temporal behavior of $\langle
\sigma_{{\rm BSM}}\rangle$ closely tracks $\sigma$, as shown by the
black curves in Fig. \ref{stressvstime}. The rapid fluctuations in
stress are accurately captured by BSM, with excellent temporal
resolution.  The uncertainty in $\langle
\sigma_{{\rm BSM}}\rangle$ arises primarily from the $\sim$ 10-20\% uncertainty in the PDMS thickness and modulus (see methods).  The calculated boundary stresses are linearly related to both of these quantities.

The real power of BSM is its ability to reveal the spatio-temporal
dynamics of stress heterogeneities that cannot be resolved with
conventional rheology. For example, the high stress event captured 
in Fig. \ref{stressvstime}D is a spatial average of the BSM data seen
in Fig. \ref{42event} and in supplementary movie 1. 
Figure \,\ref{42event}A shows the average stress vs. time (black) along with
the normal force $F_N$  reported by the rheometer divided by the area of the plate, $A_p$ (blue).
The measurement of $F_N$ is noisy, but shows large positive values
that rise and fall with the shear stress, indicating that the high
stress events are associated with strong dilatancy.
Figure \ref{42event}B shows a snapshot of the stress map 
0.42 seconds after the beginning of the event (point {\em iii} in
Fig. \ref{42event}A), showing two localized regions of high stress.
In Fig. \ref{42event}C, we show the time evolution of the stress by
''unwrapping'' the circular domain onto a rectilinear $r-\theta$ plane. The high
stress region initially appears closer to the edge of the plate, where
shear rate is maximum (Fig. \ref{42event}C, {\em i}).  As the
stress rapidly increases, the high stress region splits in two, with
one region moving in flow direction and another in the opposite
direction ({\em ii})).  The  regions grow and
periodically bifurcate, leaving behind band-like structures extended
in vorticity direction ({\em iii}), with stresses below
that of the main `parent' region, but still significantly higher than
the background stress. The high stress region which moves backwards
carries higher stress than one moving forward. When the forward and
backward moving regions collide (between $iii$ and $iv$), the stress
rapidly decreases, but the weaker band-like structures persist for
some time (Fig. \ref{42event}C {\em iv-vi}, note change in colormap).  By
comparing the movement of the high stress regions with the top plate
we see that the region moving in the flow direction is actually
rotating slightly faster than the top plate, while the weaker
band-like structures that originated from the region moving in the
flow direction rotate with approximately the same velocity as the top
plate. The bands originating from the backwards-moving regions
remain approximately motionless. Interestingly, after the collision of
the parent blobs, the motionless bands disappear, leaving only bands
moving with the top plate.  This behavior is most clearly seen in
supplementary movie 1. The movie is cropped to include only the time surrounding the two high stress events visible in Fig. \ref{stressvstime}A,
and the solid line represents the motion of the top plate (at this shear rate, one full rotation of the top plate takes approximately 3 seconds).

This same temporal evolution is observed in both high stress events
occurring at $\dot{\gamma} = 27.3$ s$^{-1}$ (see supplementary movie 1).  At $\dot{\gamma} = 29.25$ s$^{-1}$, we observe multiple events
whose initial dynamics is identical to those at $\dot{\gamma} = 27.3$
s$^{-1}$ (Fig.  \ref{45event}, supplementary movie 2). A single localized region of high stress that quickly splits
into two, one moving in the flow direction slightly faster than the
rotating top plate with the other moving opposite the flow direction,
with a sharp reduction in stress at the point when the
counter-propagating regions collide, and remnant band-like structures
that initially rotate with either the top plate or are approximately
motionless (Fig.  \ref{45event}$vi-xi$, circled, and supplementary movies). 
In contrast to the
behavior at smaller $\dot{\gamma}$, the localized regions of high
stress sometimes move in the vorticity direction, and occasionally show
more complex dynamics (supplementary movie 2, esp. the last few
events).

\begin{figure*}
\includegraphics[width=0.7\textwidth]{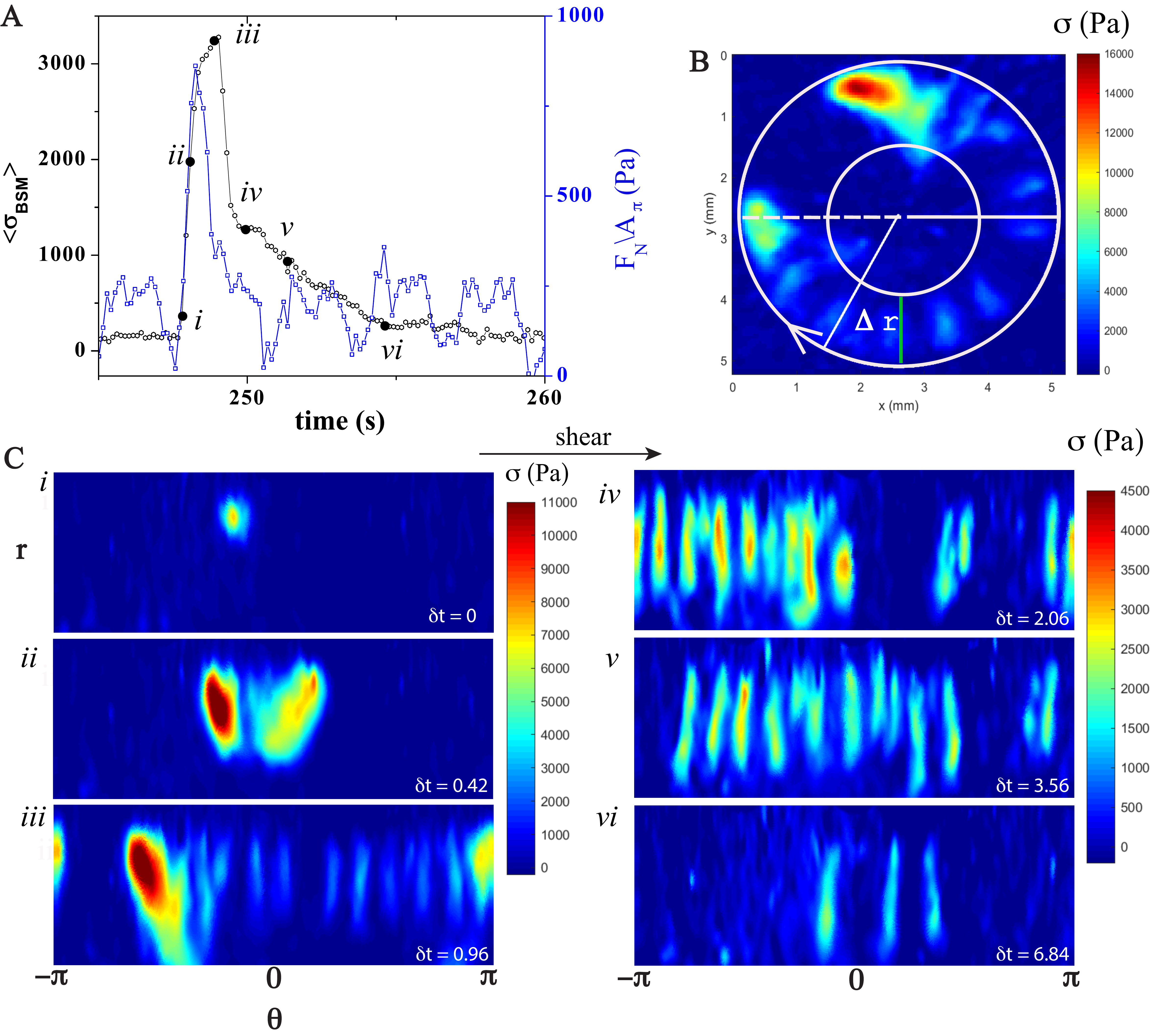}
\caption{Spatiotemporal dynamics of an individual high stress event.  A) Rheometer stress and normal force divided by the plate area for the event shown in Fig. \ref{stressvstime}D ($\dot{\gamma}$ = 27.3 s$^{-1}$, rotation period T=3 s).
B) Snapshot of boundary stress map at the peak stress (point $iii$ in A).   The rheometer plate boundary is demarcated by the white outer
  circle. C) Stress map at different time points unwrapped into  rectilinear $r-\theta$ coordinates, where $\theta$ is measured from the horizontal (solid line in B) in the flow (clockwise) direction.  The event initiates at $t_i = 247.52$ (point $i$ in A) with a localized region of high stress slightly above the horizontal (negative $\theta$), near the outer edge (large $r$).  The time elapsed from initiation, $\delta t= t-t_i$, is shown on each panel. 
 The high stress region splits into two, one moving in the  flow direction (increasing $\theta$) and the other in the opposite direction. Note that the stress scale is reduced  for ($iv-vi$), after the collision of the counter-propagating high stress regions.   }
\label{42event}
\end{figure*}

\begin{figure*}
\includegraphics[width=1\textwidth]{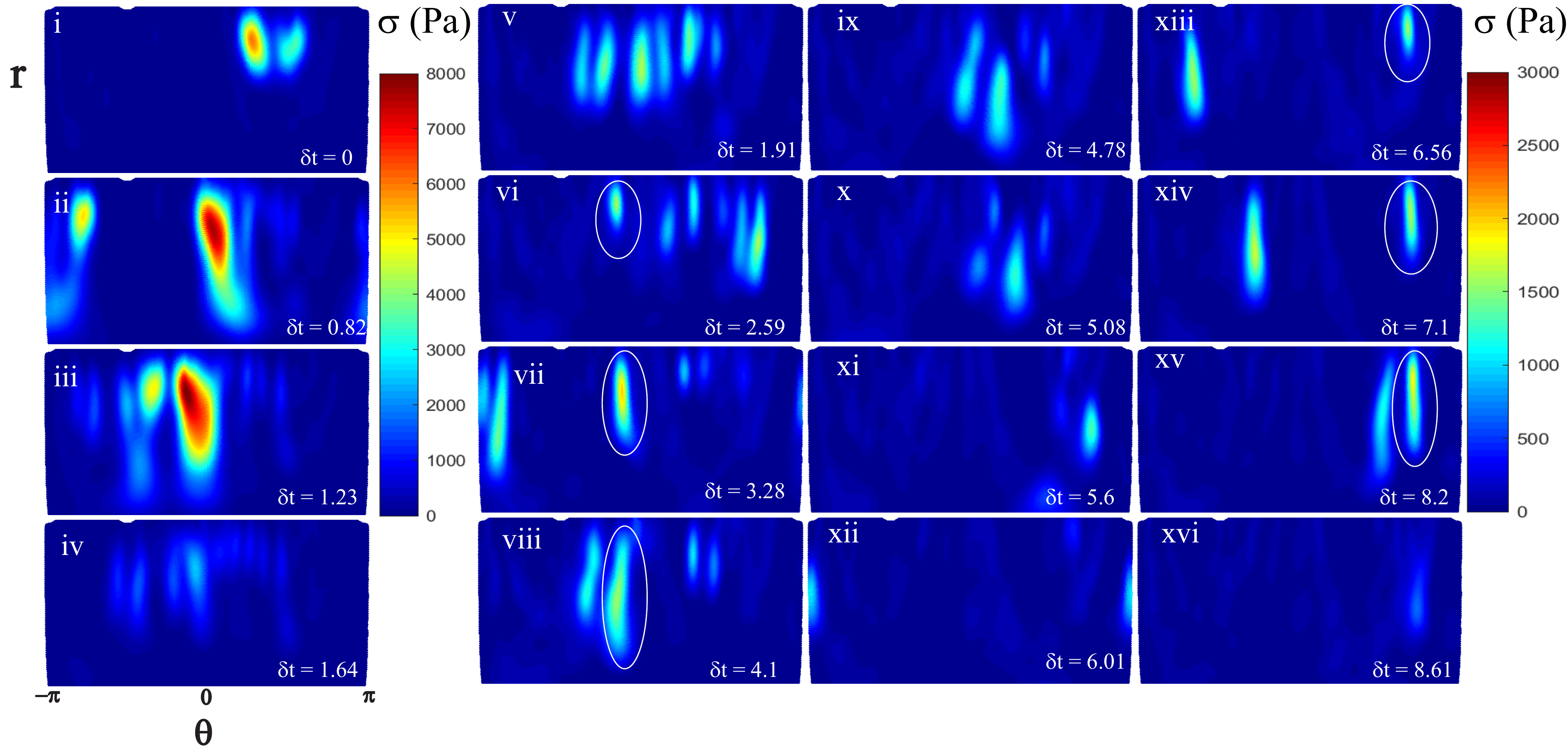}
\caption{Spatiotemporal dynamics of a high stress event at
  $\dot{\gamma}$ = 29.25 s$^{-1}$ shown in the $r-\theta$ plane, as in Fig.\ref{42event}.
 The white ellipses highlight non-propagating regions. Note that the stress scale is reduced for ($v-xvi$). }
  

\label{45event}
\end{figure*}


At still higher shear rates, multiple high stress events overlap, and
the rapidly fluctuating average stress
(e.g. Fig. \ref{stressvstime}C,F) is reflective of nearly continual,
complex spatiotemporal fluctuations in the boundary stress, with a mixture of high stress regions that propagate in the flow direction, opposite the flow direction, and in the vorticity direction (supplemental movie 3).

\subsection{Constant Applied Stress}

\begin{figure*}
\includegraphics[width=1\textwidth]{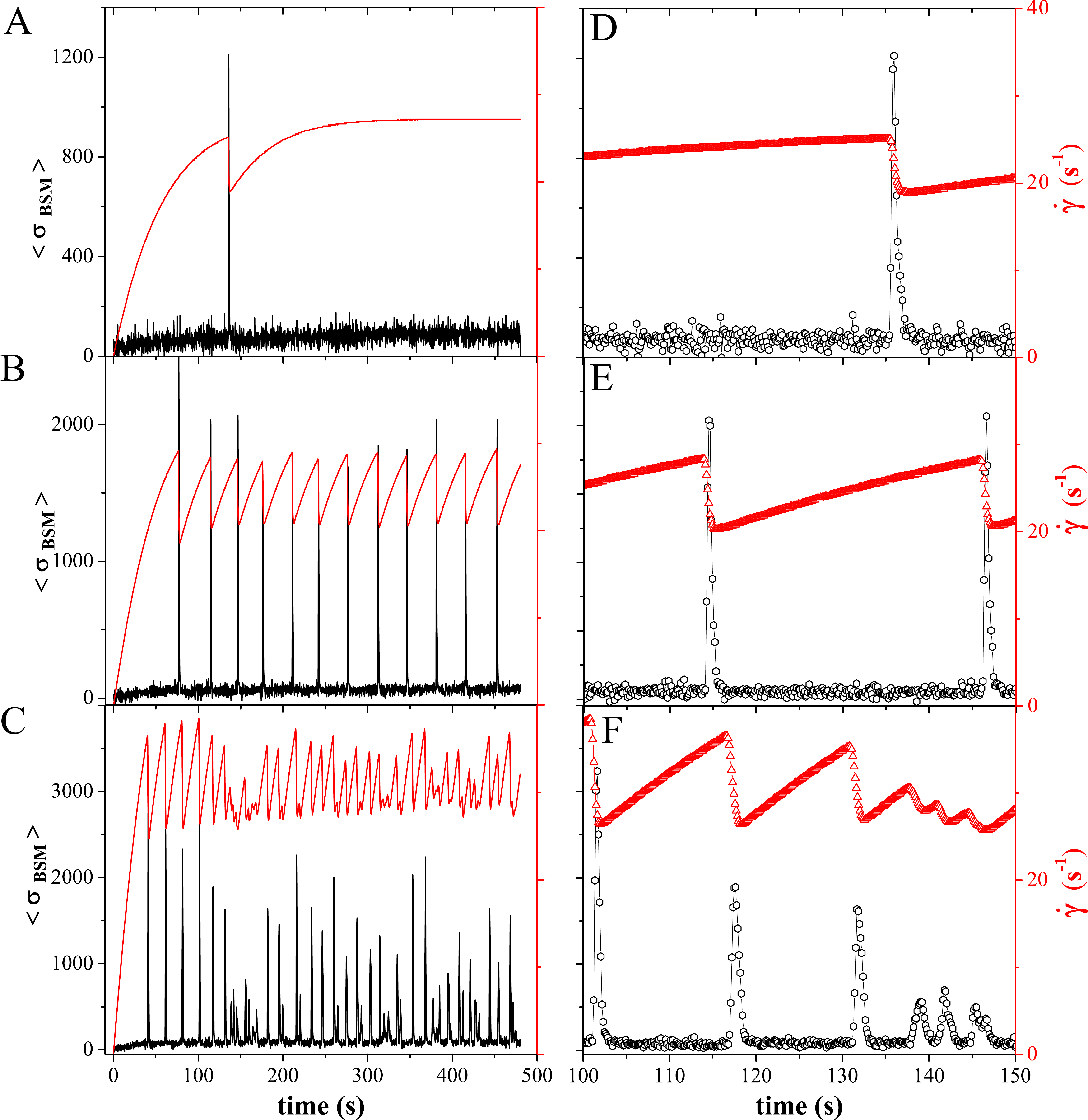}
\caption{  Shear rate (red) and average
  boundary stress (black) vs. time at  constant applied  stress $\sigma$ =   A,D)  126 Pa, B,E) 180 Pa, and C,F) 324 Pa.   } 
\label{constantstress}
\end{figure*}

In this measurement mode, the rheometer applies a constant torque to the tool, and therefore reports a constant average stress.    For a fluid with linear rheology initially at rest, the application of a fixed torque will result in an exponential approach to the steady state shear rate, $\dot{\gamma}=\sigma/\eta$, with a time constant that depends on the inertia of the tool and the fluid viscosity, as well as geometric factors.    This is indeed what we observe for low applied stress (data not shown).  However, once the stress is large enough to generate shear rates above $\dot{\gamma}_c$, the minimum shear rate at which high stress events are observed in the constant shear rate measurements, the increasing shear rate is interrupted by an abrupt drop in shear rate to a value well below $\dot{\gamma}_c$.  The drop is then followed by a resumption of the slow increase in $\dot{\gamma}$ (Fig. \ref{constantstress}A,D, red points, $\sigma$= 126 Pa applied stress).  At higher applied stress, the rotation rate increases more rapidly, while the abrupt drops occur at slightly higher shear rates, and the time between drops decreases (Fig. \ref{constantstress}B,E, $\sigma$= 180 Pa).   At still higher stress, the behavior becomes less regular, but the basic features persist (Fig. \ref{constantstress}C,F, $\sigma$= 324 Pa).  This pattern of behavior is consistent with prior observations of suspensions in the DST regime \cite{Larsen}.  
 
As with the measurements at steady shear rate, we can compare the
(constant) stress reported by the rheometer with the average boundary
stress, $\langle \sigma_{{\rm BSM}}\rangle $, calculated directly from
the deformation of the elastic substrate.  During the period of slow
acceleration, $\langle \sigma_{{\rm BSM}}\rangle$ is essentially below our measurement
resolution, but during the abrupt drops in shear rate we observe
dramatic jumps in boundary stress (Fig \ref{constantstress}, black points), similar to those observed in both $\sigma$ and
$\langle\sigma_{{\rm BSM}}\rangle$ for the constant shear rate measurement
(Fig. \ref{stressvstime}).  Note that $\sigma$ reported by the rheometer is constant in this measurement mode.   This highlights the fact that 
$\sigma$, calculated from the torque applied to the
tool, is {\em not} the average stress at the boundary of the
suspension when the rotation rate varies, because of the inertia of
the tool.  Thus the dramatic stress fluctuations revealed by BSM are
missed in standard bulk rheology, although they can be inferred from
the changes in shear rate with quantitative modeling that incorporates
the inertia of the rheometer tool \cite{Larsen, Bossis, Poon_PRL_2019}.

\begin{figure*}
\includegraphics[width=0.6\textwidth]{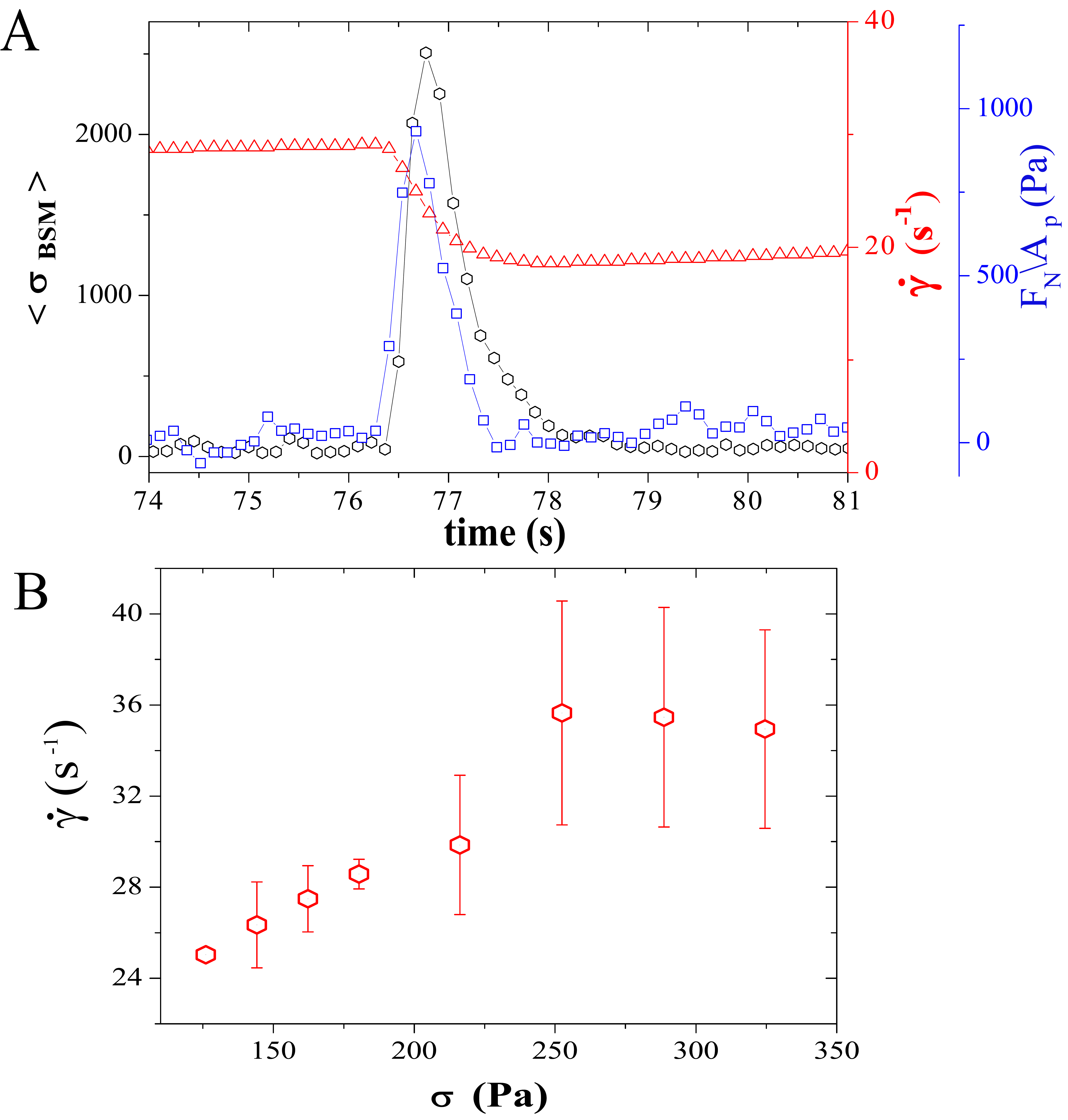}
\caption{A) Average boundary stress (black circles),
  $\dot{\gamma}$ (red triangles) and normal force  (blue squares) for one event at $\sigma$ = 180 Pa. B) Shear
  rate at which high stress events nucleate as a function of applied
  stress. }
\label{constantstressevents}
\end{figure*}

Figure \ref{constantstressevents}A shows $\sigma_{{\rm BSM}}$ (black
circles) for a single event, along with $\dot{\gamma}$ (red triangles) and the 
normal force ($F_N$, blue squares), at an applied
stress of $\sigma$ = 180 Pa.  The abrupt change in $\sigma_{BSM}$ is
also present in $F_N$, as with the events at constant shear rate,
and consistent with previous reports \cite{Lootens_2}. Note that the sign of $F_{N}$ depends on the relative contribution of lubrication and frictional forces which can result in a negative  F$_{N}$ even in discontinuous shear thickening \cite{Nakanishi_1, Bonn_3,  Nott}.    The
events typically have a very sharp onset, allowing us to define an
onset shear rate for each event.  Figure \ref{constantstressevents}B
shows the average of the onset shear rates found for each applied
shear stress.  For $\sigma$ = 126 Pa, only one event is observed; for
all higher values of $\sigma$, the error bars represent the standard
deviation of the onset shear rates.  At the lower $\sigma$, the
events are well separated (Fig. \ref{constantstress}A,B), while at
higher stresses the events begin to overlap, which may explain the
change in behavior above 210 Pa.  
 
\begin{figure*}
\includegraphics[width=1\textwidth]{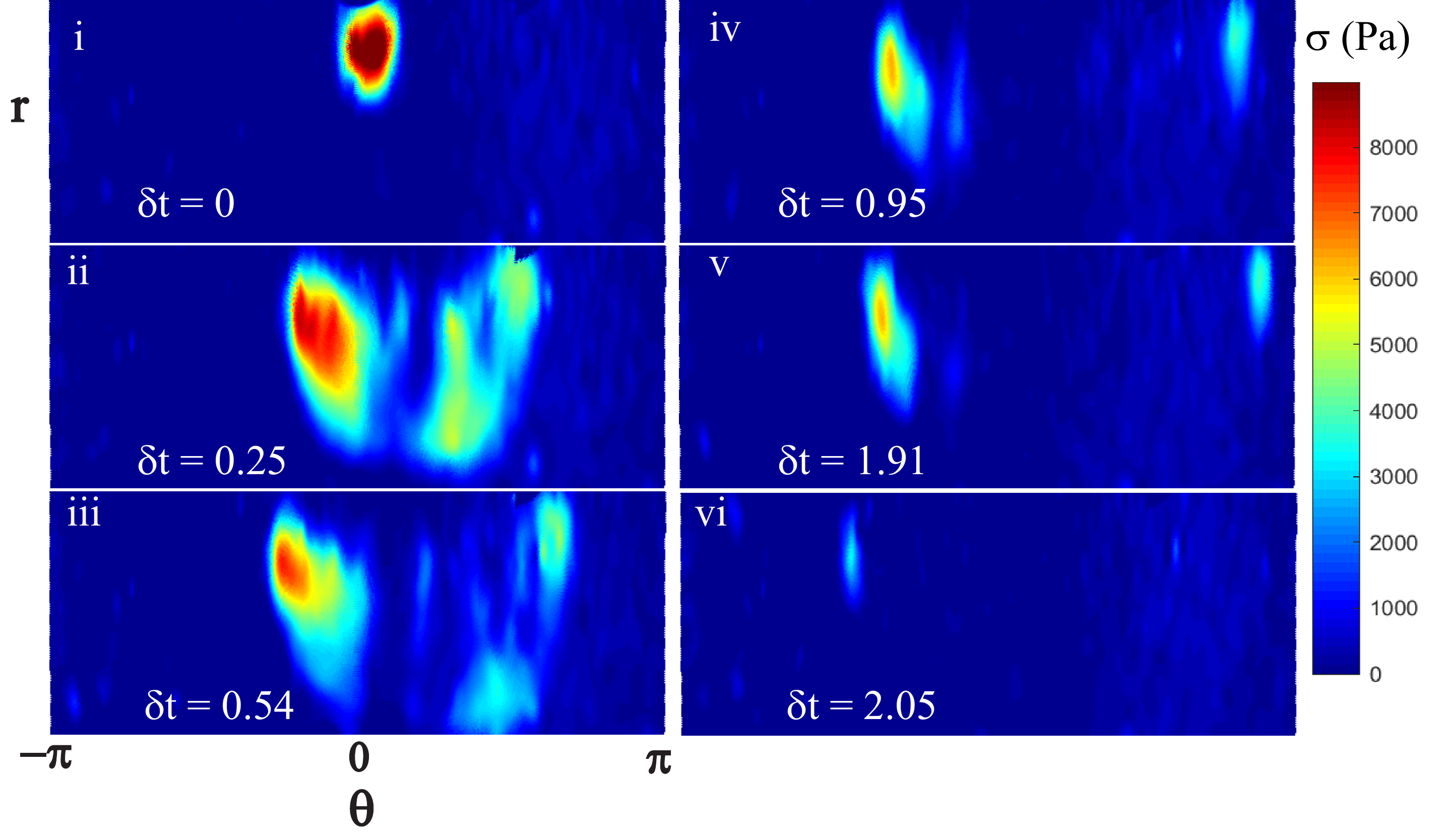}
\caption{ Spatiotemporal dynamics of a high stress event at constant applied stress ($\sigma$) = 180 Pa hown in the $r-\theta$ plane, as in Fig.\ref{42event}.}
\label{50Paevent}
\end{figure*}

The typical spatiotemporal dynamics of one of these events is displayed in Fig. \ref{50Paevent} (see also supplementary movie 4).   The initial evolution is similar to the first phases of the events seen at constant shear rate (Fig. \ref{42event}),  (i) a rapid initiation of a large localized stress close to the edge of the tool where the shear rate is largest, associated with a rapid, large increase in normal stress, followed closely by (ii) a bifurcation into two  regions, one  moving in flow direction and another in the opposite direction.
However, unlike the measurements at constant shear rate, the high stress regions quickly decay, presumably as a result of the sudden drop in shear rate to a value  $\dot{\gamma}<\dot{\gamma}_c$, so do not survive long enough to collide after collectively completing a trip around the circle. 
 
\section{Discussion}
\label{discussion}

\subsection{Model for Dynamics of High Stress Phases:}  
The striking behavior revealed by boundary stress microscopy 
suggests four distinct stages for the high stress events observed at constant applied shear rates slightly above $\dot\gamma _c$: ($i$) rapid
initiation of a large localized stress, associated with a rapid, large
increase in normal stress, followed closely by ($ii$) a bifurcation
into two high stress regions, one of which is anchored to the
bottom plate, moving slowly upstream, the other of which is anchored
to the top plate, moving slowly downstream relative to the top plate. ($iii$) During the
migration, the two regions (mostly the one anchored to the top plate) leave
behind band-like regions of high stress extended in the vorticity
direction and ($iv$) a
dramatic drop in stress when the counter-propagating regions collide.

\begin{figure*}
\includegraphics[width=0.7\textwidth]{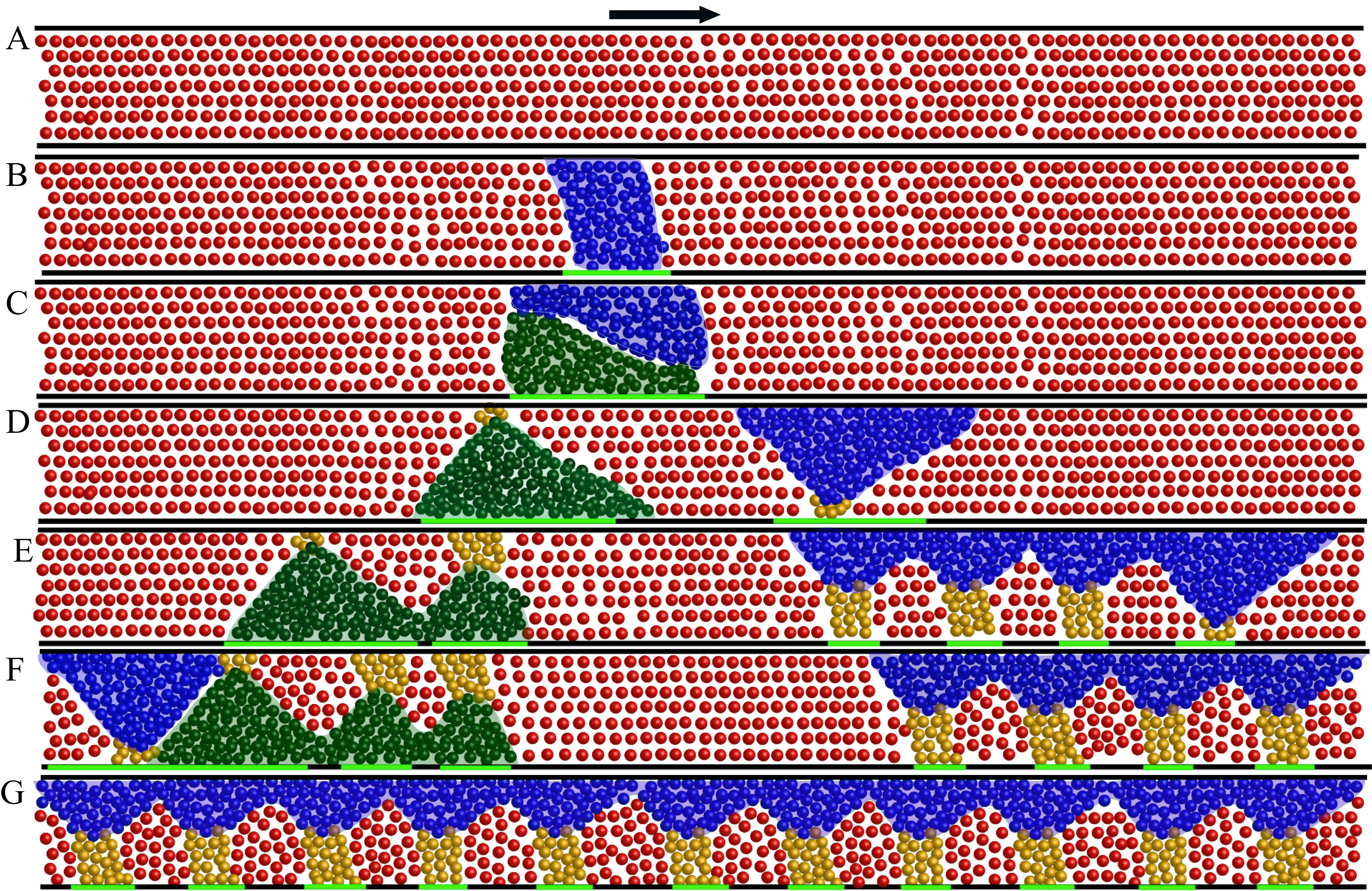}
\caption{Schematic of evolution of the solid-like phases (SLPs) inferred from the BSM measurements.  (A) The relatively homogeneous, smoothly sheared suspension is interrupted by a  gap-spanning SLP (B, blue particles), where the dilatant pressure creates direct contact between the particles and the boundaries and a resulting region of high boundary stress (green).  The motion of the rheometer tool causes the SLP to fracture into two regions (C). One  region remains anchored to the bottom plate (dark green spheres)  and moves slowly backwards (D, E), while the upstream side moves in the direction of the flow (D, E).  The increased shear rate between the SLPs and the rheometer plates causes the  flowing suspension transitions to a high viscosity (frictional) fluid branch (gold particles).
Both regions accumulate particles on their outer sides and split off smaller band-like regions of SLPs on their inner sides.  The two large SLPs collide (F) and annihilate, leaving residual
band-like structures that move with the top plate (G), which continue to generate regions of relatively large boundary stress (green).}
\label{schematic}
\end{figure*}

Phase (i) is consistent with the formation of a localized shear jammed
region, or solid-like phase (SLP) that, because of the dilatant pressure resulting from
frictional contacts, makes strong physical contact with the
boundaries, penetrating the thin fluid layer that normally exists
between the suspended particles and the boundaries.  
This  is
represented schematically in Fig. \ref{schematic}, where the
relatively homogeneous, smoothly sheared suspension (Fig. \ref{schematic}A) is interrupted by a region with a fluctuation to a higher
density, where an increase in local shear stress produces an increase
in frictional contacts, thereby creating a gap-spanning region of increased frictional contacts (Fig. \ref{schematic} B, blue particles).   Alternatively, the initial high stress region might consist of SLPs on the top and bottom plate, with a flowing fluid phase in between.
In either case the increased dilatant pressure will create
direct contact between the particles and the rheometer tool, and high
stress detected by BSM (show schematically by the green region on the
bottom boundary). The rheometer stress will increase to the whatever
value is necessary to keep the tool rotating, so the bifurcation into
two SLPs (phase ii) arises from a fracture of the jammed
solid, with one SLP staying approximately anchored to the bottom
plate, and the other anchored to the top plate
(Fig. \ref{schematic}C). 

Because the shear rate in the SLPs is zero (or close to it), the shear rate in the suspension traversing between the SLPs and the rheometer plates will be higher than the average shear rate.  We hypothesize that this increased shear rate will drive a transition to a high viscosity (frictional) fluid branch (Fig. \ref{schematic}C, gold particles). The co-existence of two flowing phases of very different viscosities is consistent with our prior BSM measurements of suspensions in CST \cite{vikram}.  The presence of a high viscosity phase helps explain the very large stresses observed in BSM (nearly two orders of magnitude higher than the background boundary stress).

As the top plate rotates, the SLP anchored to the bottom plate (black
spheres) moves slowly backwards, perhaps as a result of a densification of the
upstream side caused by an accumulation of particles from the flowing
suspension.  The reverse will happen with the SLP anchored to the top
plate (blue spheres), as it is swept through the suspension at a
velocity that exceeds the local affine flow speed.  This is shown
schematically by the additional particles added to the jammed region
(Fig. \ref{schematic}D). Because the flowing suspension transitions to the high viscosity fluid phase as it transits the reduced gap between the SLP and the bottom boundary, the SLPs attached to the top
plate continue to show up as large boundary stresses on the bottom
plate.  It is unclear why the growing SLPs would
divide and leave behind band-like jammed regions, or what sets the characteristic spacing between the bands, but the bands themselves remain approximately motionless relative to the plates
(smaller black and blue regions).  When the two large `parent' SLPs collide
(iv, Fig. \ref{schematic}F), the interaction presumably disrupts the
network of frictional contacts, resulting in a return to a mostly
freely flowing low viscosity suspension, except for the residual
band-like structures that are moving with the top plate (Fig. \ref{schematic}G). 

A similar picture applies to the  events observed at constant applied stress (Fig. \ref{50Paevent}), except that the nucleation of the gap-spanning SLP is followed quickly by a drop in shear rate.  We hypothesize that the flowing suspension will no longer transition to the high viscosity fluid phase, and the shear stress will be insufficient to maintain the shear-jammed state, so the SLPs will erode.

\subsection{Connection with Previous Observations:}
These results are consistent with previous observations of
fluctuations in DST \cite{Boersma, Larsen, Bossis, Nakanishi_1, Nakanishi_2, vikram_2,  Poon_JoR, Poon_PRL_2019}, but the power of BSM to resolve the
spatiotemporal dynamics of heterogenous stresses reveals a complex and
subtle evolution of those fluctuations.  Larsen et al. \cite{Larsen}
observed shear rate fluctuations similar to those in
Fig. \ref{constantstress} with a connection to increases in normal
stress, and speculated that these were due to transient periods of
dilatant, solid-like behavior, with the dilatancy responsible for a
breakdown of the slip layer between the flowing particles and the
boundaries.  Hermes et al. \cite{Poon_JoR} observed that, in
suspensions of corn starch particles in DST, rapid decreases in shear
rate were accompanied by local deformations of the air-sample
interface at the edge of the rheometer tool, and that the deformations
sometimes appear static and sometimes move opposite to the direction of
the flow.  The surface deformations are also consistent with the
existence of substantial dilatant pressure.

The interplay of dilatant pressure, surface deformations, and the presence or absence of a fluid layer between the particles and the rheometer plates suggests that boundary conditions play an important role in determining the spatiotemporal evolution of the high stress regions.  While we cannot directly measure local normal stresses, they are likely of similar magnitude to the shear stress, which will produce deformations of the incompressible but compliant PDMS layer that will transiently change the gap by a few microns (i.e. a few percent). The fact that we and others see similar intermittent fluctuations in the system-averaged stress in the absence of a compliant boundary suggest that the basic phenomena is not dependent on those deformations, but the detailed evolution may be different with differing amounts of boundary compliance.

We previously reported the results of BSM measurements in the
continuous shear thickening regime \cite{vikram}, where we found
 localized regions of high stress appearing intermittently
at stresses above the critical stress. However, unlike the high stress
regions reported here, the high stress regions in CST propagate in the flow direction with an average speed of about one-half of that
of the top plate.  We interpret these high stresses as
arising from a high viscosity fluid phase, with approximately affine
shear throughout, as opposed to the localized regions of jammed solid phases
that we infer here from BSM measurements in DST.  This picture is consistent with
the prediction of the mean field model of Wyart and Cates of a
transition from low to high viscosity fluid phases at moderately high
concentration, and a transition from fluid to jammed solid phases at
still higher concentration \cite{Cates_1}.  In our previous report we
also showed that close to DST, some high stress regions do not
propagate, and that the tracer particles in the suspension tens of
microns above the bottom plate intermittently become motionless,
indicating that a localized portion in the suspension jammed into a
fully solid phase \cite{vikram}, consistent with the behavior shown in
Fig. \ref{schematic}.  Those measurements were performed at high
magnification, so we were not able to resolve the rapidly moving
events that would correspond to solid regions anchored to the top
plate.

These results reveal some of the complexity of the spatiotemporal dynamics of shear thickening suspensions,  which, in general, will  depend on the type of suspension, the boundary conditions, and the measurement geometry.   In particular, curvature and shear rate gradient effects may be significant in the small tool parallel plate geometry used in this study.  More generally, a full understanding of these phenomena will require continuum models that can move beyond mean-field and include descriptions of non-affine flow and associated density fluctuations, the coupling between flow, density, and stress, and the connection between dilatant pressure and suspension confinement.

\section{Supplementary Material}
See supplementary material for temporal evolution of stress using a custom  rheometer tool in absence of PDMS film and movie captions.

\section{Acknowledgments}

We  thank  Leon Der for fabrication of the custom rheometer tool and Emanuela  Del  Gado and Peter Olmsted for  helpful  discussions.  This  work  was supported  by  NSF grant DMR-1809890, and NSF grant PHY-1748958 through the KITP program on the Physics of Dense Suspensions.  J.S.U.  is  supported,  in  part,  by  the  Georgetown  Interdisciplinary  Chair  in Science Fund.

\newpage

\section{Supplementary Information}
\makeatletter
\renewcommand{\fnum@figure}{\figurename~S1}
\makeatother
\begin{figure*}
	\includegraphics[width=0.81\textwidth]{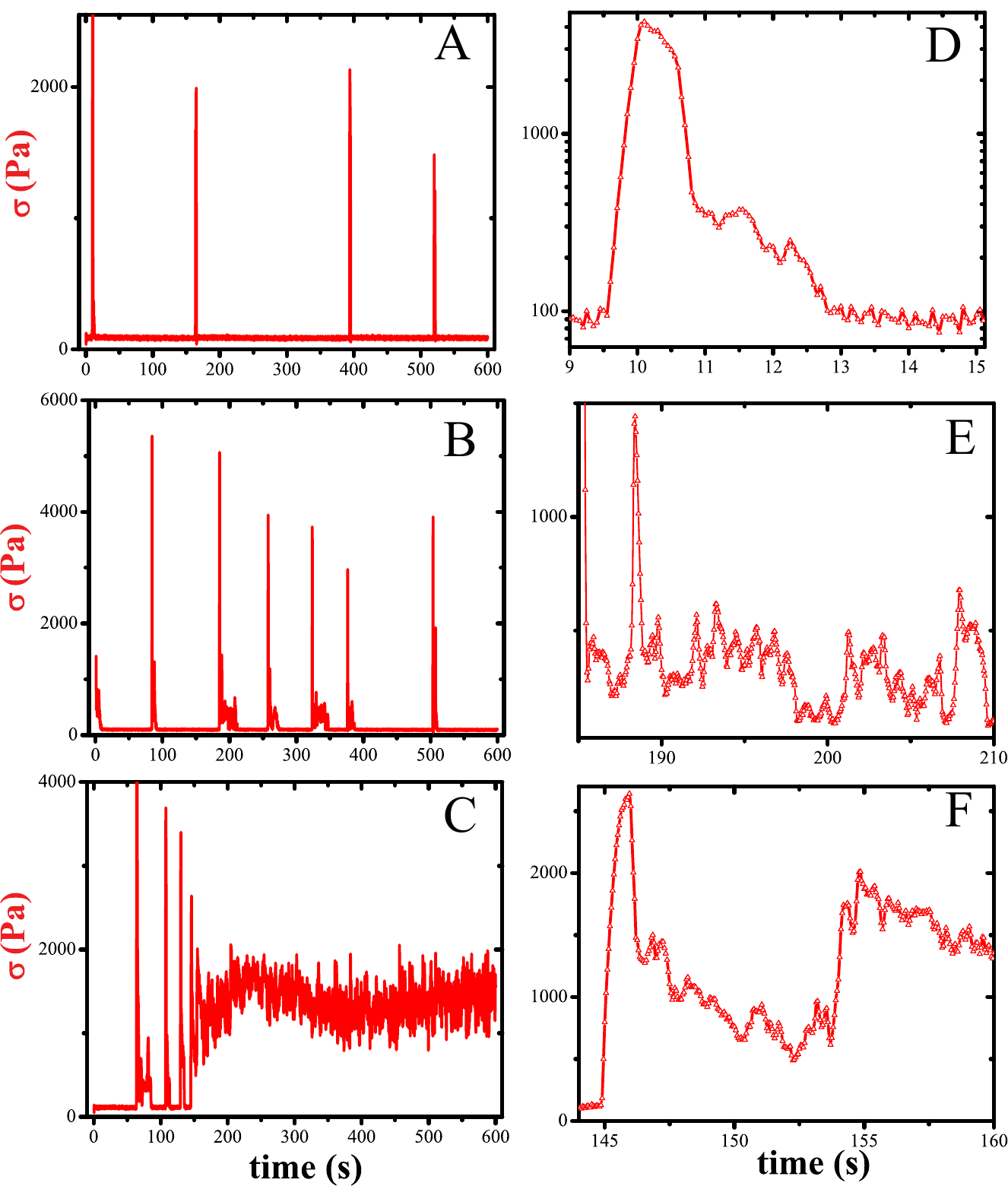}
	\caption{ Temporal evolution of stress ($\sigma$) for different applied shear rates ($\dot{\gamma}$) A, D) 29.3 s$^{-1}$, B, E) 32.5 s$^{-1}$ and C, F) 37.4 s$^{-1}$ for $\phi$ = 0.56 using the custom rheometer tool as described in the Methods, except without a PDMS film on the bottom plate.}
	\label{supp_1}
\end{figure*}

\section{Supplementary Video Captions}

\noindent\textbf{Supplementary Movie 1:}  Spatiotemporal dynamics of high stress events revealed by boundary stress measurements   at a shear rate ($\dot{\gamma}$) = 27.3 s$^{-1}$ for  $\phi$ = 0.56. The movie is cropped to include only the time surrounding the two high stress events are visible in Fig. 2A (main text). The solid line and circle represents the motion and edge of the top plate, respectively.

\noindent\textbf{Supplementary Movie 2:} Spatiotemporal dynamics of high stress events revealed by boundary stress measurements  at a shear rate ($\dot{\gamma}$) = 29.25 s$^{-1}$ for  $\phi$ = 0.56. The movie is cropped to include only  time surrounding for three high stress events.  The solid line and circle represents the motion and edge of the top plate, respectively.

\noindent\textbf{Supplementary Movie 3:} Spatiotemporal dynamics of boundary stresses at shear rate ($\dot{\gamma}$) = 39 s$^{-1}$ for  $\phi$ = 0.56. Partial movie is shown (first 100 seconds).  The solid line and circle represents the motion and edge of the top plate, respectively.

\noindent\textbf{Supplementary Movie 4:} Spatiotemporal dynamics of boundary stresses at constant applied stress  ($\sigma$) at  = 180 Pa, for  $\phi$ = 0.56. The circle represents edge of the top plate.

\end{document}